\documentclass[a4paper,prl,twocolumn,showpacs,superscriptaddress,floatfix,nofootinbib]{revtex4-1}
\usepackage{times}
\usepackage{amssymb,amsmath}

\newcommand{\D}{\mathrm{D}}
\newcommand{\di}{\mathrm{d}}

\newcommand{\vph}{\varphi}
\newcommand{\vth}{\vartheta}

\usepackage{graphicx}
\usepackage{dcolumn}
\usepackage{bm}
\usepackage{xcolor}
\definecolor{navy}{RGB}{0,0,150}
\usepackage[colorlinks=true,allcolors=navy]{hyperref}
\usepackage{fancyhdr}
\usepackage{orcidlink}
\begin{document}
\preprint{APS/123-QED}
\title{Analytic solutions for the motion of spinning particles near spherically symmetric black holes and exotic compact objects}
\author{Vojt{\v e}ch Witzany\,\orcidlink{0000-0002-9209-5355}}
\email{vojtech.witzany@matfyz.cuni.cz}
\affiliation{%
Institute of Theoretical Physics, Faculty of Mathematics and Physics, Charles University, CZ-180 00 Prague, Czech Republic}
\author{Gabriel Andres Piovano\,\orcidlink{0000-0003-1782-6813}}
\email{gabriel.piovano@ucd.ie}
\affiliation{%
School of Mathematics and Statistics, University College Dublin, Belfield, Dublin 4, Ireland}
\date{\today}
\begin{abstract}
Rapidly rotating bodies moving in curved space-time experience the so-called spin-curvature force, which becomes important for the motion of compact objects in gravitational-wave inspirals. As a first approximation, this effect is captured in the motion of a spinning test particle. We solve the equations motion of a spinning particle to leading order in spin in arbitrary static and spherically symmetric space-times in terms of one-dimensional closed-form integrals. This solves the problem and proves its integrability in a wide range of modified gravities and near exotic compact objects. Then, by specializing to the case of bound orbits in Schwarzschild space-time, we demonstrate how to express the solution in the form of Jacobi elliptic functions. 
\end{abstract}
\maketitle

\noindent{\em Introduction.}--
After a track record of spectacular successes in their first three observing runs \cite{LIGOScientific:2018mvr,LIGOScientific:2020ibl,KAGRA:2021vkt}, in  May 2023 the LIGO and Virgo instruments were joined for the first time by the Japanese detector KAGRA to start the fourth run of observing coalescing compact objects through gravitational waves \cite{KAGRA:2013rdx}. The so-called LVK collaboration is expected to detect more events in this run than have been amassed to date. Next-generation detectors on Earth and in space promise to multiply the sensitivity and reach of the detectors even further \cite{Kalogera:2021bya,amaro2017}. As the number of events grow, we will observe binaries with different mass ranges, mass ratios, and dynamical setups and, consequently, we will need a more faithful and complete picture of the two-body dynamics in order to detect and interpret these signals correctly. 

In particular, the space-based observatory Laser Interferometer Space Antenna (LISA)  will detect inspirals of stellar-mass compact objects (henceforth dubbed as secondary) onto supermassive black holes \cite{babak2017}. This will allow to accurately test whether the supermassive objects in the centres of galaxies are truly black holes or perhaps some exotic compact objects, or whether their gravitational field is described by Einstein gravity \cite{LISA:2022kgy}. These so-called extreme-mass-ratio inspirals (EMRIs) are best modelled in a mass-ratio expansion \cite{Barack:2018yvs}. At leading order, the two-body dynamics is approximated by the adiabatic inspiral of a test particle in the space-time of the massive compact object, where the inspiral is driven by fluxes sourced by the secondary \cite{Hughes:2021exa,Hughes:2005qb}.
Subleading corrections to the motion are collectively called post-1-adiabatic, and include the effects of second-order in the mass-ratio fluxes, conservative self-force, and the spin of the smaller companion~\cite{vandeMeent:2017bcc,Warburton:2021kwk,Wardell:2021fyy}. 
Post-1-adiabatic corrections are fundamental to model waveform suited for data analysis for LISA~\cite{babak2017,Barack:2018yly}, and the spin of the secondary plays a crucial role~\cite{Huerta:2011zi,Huerta:2011kt,Piovano:2021iwv}.
The effects of the latter are fully captured by the motion of a spinning test particle in the space-time of the massive compact object (computed to linear order to spin), and the outgoing gravitational-wave flux it sources \cite{Warburton:2017sxk,Akcay:2019bvk,Mathews:2021rod,Skoupy:2021asz,Skoupy:2022adh,Skoupy:2023lih,Piovano:2020zin,Piovano:2022ojl,Drummond:2023wqc}. Here we present an analytical solution of this motion near spherically symmetric compact objects in terms of quadratures. In particular, our analysis implies that the spinning particle motion is integrable at linear order in spin in any static, spherically symmetric space-time. We then specialize to Schwarzschild space-time and re-express the solution in terms of Jacobi elliptic integrals. Even though a number of works have treated this and similar topics previously (see, e.g., Refs \cite{Rietdijk:1992tx,Suzuki:1997by,Jefremov:2015gza,Zelenka:2019nyp,Drummond:2022efc,Drummond:2023wqc}), this is the first time an analytic, closed-form solution of generic bound motion of a spinning test particle in Schwarzschild space-time is presented.

We use the $G=c=1$ geometrized units and the (-+++) signature of the metric. Greek indices label coordinate components. The Riemann tensor is defined by $a_{\nu;\kappa\lambda} - a_{\nu;\lambda\kappa} \equiv R^\mu{}_{\nu\kappa\lambda} a_\mu$, where $;$ denotes the covariant derivative and $a_\mu$ is an arbitrary form.


\noindent{\em Spinning particle.}--
The motion of a spinning particle is given by the equations of motion \cite{mathisson1937, papapetrou1951,dixon1964}
\begin{align}
    &\frac{\D P^\mu}{\di \tau} = - \frac{1}{2} R^\mu{}_{\nu\kappa \lambda}\dot{x}^\nu S^{\kappa \lambda}\,, \\
    &\frac{\D S^{\mu\nu}}{\di \tau} = P^\mu \dot{x}^\nu - P^\nu \dot{x}^\mu \,,
\end{align}
where $P^\mu$ and $\dot{x}^\nu$ are the 4-momentum and tangent vector to the worldline of the spinning test-particle, while $S^{\mu\nu} = - S^{\nu\mu}$ is the spin tensor.
 We are only interested in the dynamics to linear order in spin $S^{\mu\nu}$ and, unless specified, all formulas are to be assumed at most $\mathcal{O}(S)$ accurate. This is justified by considering a comparably light compact object in the field of a heavy black hole, since the truncated terms can be shown to scale as higher order in the mass ratio, as discussed in the Supplemental material \cite{SupplMat}. We fix the relation of the centroid $x^\mu(\tau)$ and the momentum $P^\mu$ by the Tulczyjew-Dixon supplemental spin condition $S^{\mu\nu} P_\nu = 0$, to obtain $P^\mu = m \dot{x}^\mu + \mathcal{O}(S^2)$ where $m$ is the particle mass \cite{tulczyjew1959,dixon1964,semerak1999}. The spin tensor can then be expressed as $S^{\mu\nu} = m\varepsilon^{\mu\nu\kappa\lambda}\dot{x}_\kappa s_\lambda/2$, where $s^\mu$ is the specific spin vector and $\varepsilon^{\mu\nu\kappa\lambda}$ the Levi-Civita pseudo-tensor. The equations then reduce to \begin{align}
    &\frac{\D^2 x^\mu}{\di \tau^2} = - \frac{1}{4}{R}^\mu{}_{\nu\gamma \delta}\varepsilon^{\gamma\delta}{}_{\kappa\lambda} \dot{x}^\nu \dot{x}^\kappa s^\lambda\,, \label{eq:sacc} \\
    &\frac{\D s^{\lambda}}{\di \tau} = 0 \,. \label{eq:strans}
\end{align}


\noindent{\em Motion in static, spherically symmetric metric.}--
A general static spherically symmetric spacetime metric can be locally expressed in the form
\begin{align}
    \di s^2 = - f(r) \di t^2 + h(r) \di r^2 + r^2 \left(\di \theta^2 + \sin \theta^2 \di \phi^2\right) \,.
\end{align}
The Schwarzschild metric is contained within this class by setting $f(r), h(r)$ to $f(r) = 1/h(r) = 1-2M/r$. The Killing vectors of this metric are generators of time translations and rotations around the $x,y,z$ axes:
\begin{align}
    &\xi_{(t)} = \frac{\partial}{\partial t}\,,\\
    &\xi_{(x)} = -\sin \phi \frac{\partial}{\partial \theta} - \cos \phi \cot \theta \frac{\partial}{\partial \phi}\,,\\
    &\xi_{(y)} = \cos \phi \frac{\partial}{\partial \theta} - \sin \phi \cot \theta \frac{\partial}{\partial \phi}\,,\\
    &\xi_{(z)} = \frac{\partial}{\partial \phi}\,.
\end{align}
 Dixon \cite{dixon1970} showed that for any Killing vector the spinning particle motion has a constant of the form $C_{(\xi)} = P^\mu \xi_\mu -\xi_{\rho;\sigma} S^{\rho \sigma} /2$. In this case the Killing vectors correspond to a conserved specific spin-orbital energy and a formal angular momentum vector (we normalize each $C_{(\xi)}$ by $m$)
\begin{align}
    & \mathcal{E} = -f \dot{t} + \frac{r^2 \sin \theta f' s^\phi \dot{\theta} - s^\theta \dot{\phi}}{2 \sqrt{f h}} \,,\\
    \begin{split}
    & \mathcal{J}_x = -r^2 (\sin \phi \,\dot{\theta} + \cos \phi \cos \theta \sin \theta\, \dot{\phi}) \\
    & \phantom{\mathcal{J}_x =} + \sqrt{\frac{f}{h}} 
    \Big[
    \sin \theta \cos \phi\, h (s^t \dot{r} - s^r \dot{t}) \\
    &\phantom{\mathcal{J}_x =}  + r \sin \phi \sin \theta (s^\phi \dot{t} - s^t \dot{\phi}) \\
    &\phantom{\mathcal{J}_x =}  + r \cos \phi \cos \theta (s^t \dot{\theta} - s^\theta \dot{t})
    \Big] \,,
    \end{split} \label{eq:Jx}\\
    \begin{split}
    & \mathcal{J}_y = r^2 (\cos \phi \,\dot{\theta} - \sin \phi \cos \theta \sin \theta\, \dot{\phi}) \\
    & \phantom{\mathcal{J}_x =} + \sqrt{\frac{f}{h}} 
    \Big[
    \sin \theta \sin \phi\, h (s^t \dot{r} - s^r \dot{t}) \\
    &\phantom{\mathcal{J}_x =}  + r \cos \phi \sin \theta (s^t \dot{\phi} - s^\phi \dot{t}) \\
    &\phantom{\mathcal{J}_x =}  + r \sin \phi \cos \theta (s^t \dot{\theta} - s^\theta \dot{t})
    \Big] \,,
    \end{split} \label{eq:Jy}\\
    \begin{split}
    & \mathcal{J}_z = r^2 \sin^2\! \theta \, \dot{\phi} \\
    & \phantom{\mathcal{J}_x =} +  \sqrt{\frac{f}{h}} 
    \Big[
    \cos \theta \, h (s^t \dot{r} - s^r \dot{t}) \\
    &\phantom{\mathcal{J}_x =}  + r \sin \theta (s^\theta \dot{t}-s^t \dot{\theta})
    \Big] \,.
    \end{split} \label{eq:Jz}
\end{align}
The linearisation in spin means that these integrals of motion are conserved up to $\mathcal{O}(s^2)$. Another integral of motion is obtained by noticing that the following vector is parallel-transported along geodesics in general spherically symmetric static space-times:
\begin{align}
    l =  \frac{r \dot{\theta}}{\sin \theta} \frac{\partial}{\partial \phi} - r \sin\theta \dot{\phi}\frac{\partial}{\partial \theta}\,,
\end{align}
and for spinning particles $D l^\mu/\di \tau = \mathcal{O}(s)$. As a result, the aligned component of the spin vector is an approximate constant of motion
\begin{align}
    s_{\parallel} \equiv \frac{l^\mu s_\mu}{\sqrt{l^\nu l_\nu}}\,,\;\frac{\di s_\parallel}{\di \tau} = 0+\mathcal{O}(s^2) \,.
\end{align}


\noindent{\em Angular momentum aligned coordinates.}--
For general initial conditions, the formal angular-momentum vector $\vec{\mathcal{J}} = (\mathcal{J}_x,\mathcal{J}_y,\mathcal{J}_z)$ keeps pointing into a constant direction. Additionally, the orbital plane is always almost orthogonal to this vector up to $\mathcal{O}(s)$ corrections.
Hence, without loss of generality, we can rotate into a new coordinate system $\theta,\phi \to \vth,\vph$ such that its axis of $\vph$-rotation ($\vth = 0, \pi$) points in the direction of $\vec{\mathcal{J}}$. When defined with respect to this new system, we obtain $\mathcal{J}_x' = \mathcal{J}_y' = 0, \mathcal{J}_z' = \sqrt{\mathcal{J}_x^2+\mathcal{J}_y^2+\mathcal{J}_z^2} \equiv\mathcal{J}$, and the position of the particle will fulfil $\vth(\tau) = \pi/2 + \delta \vth(\tau)$, where
\begin{align}
    \delta \vth = \frac{\sqrt{f h} (s^r \dot{t} - s^t \dot{r})}{r^2 \dot{\vph}} + \mathcal{O}(s^2)\,. \label{eq:dvth}
\end{align}
In other words, the $\vth$-motion of the particle is automatically expressed in terms of the other variables. This transformation was already implicitly used in the numerical studies of Refs \cite{Suzuki:1997by,zelenka2019}, even though the authors did not realise it also allows to analytically solve for the $\vth$ degree of freedom at $\mathcal{O}(s)$. More details of the computation can be found in the Supplemental material. 

The equations of motion for the other orbital variables $r(\tau),t(\tau),\vph(\tau)$ can now be expressed in first-order form up to $\mathcal{O}(s^2)$
\begin{align}
    & \dot{r}^2 = \frac{1}{h} \left( -1 + \frac{\mathcal{E}^2}{f} - \frac{\mathcal{J}^2}{r^2}\right) + \frac{s_\parallel \mathcal{E} \mathcal{J} (2 f - r f')}{ (f h)^{3/2} r^2}\,, \label{eq:dotr}\\
    & \dot{t} = \frac{\mathcal{E}}{f} + \frac{s_\parallel \mathcal{J} f'}{2 f r \sqrt{fh}}\,,\\
    & \dot{\vph} = \frac{\mathcal{J}}{r^2} + \frac{s_\parallel \mathcal{E} }{ r^2 \sqrt{fh}}\,.
\end{align}


\noindent{\em Solving parallel transport.}--
As for the spin degree of freedom, we need to solve the parallel transport equation at leading (geodesic) order. To do so, we construct a parallel-transported tetrad $e^\mu_{(A)}, A=0,...,3$ inspired by Marck \cite{marck}. 
We start with the zeroth and third leg (component order $t,r,\vph,\vth$) 
\begin{align}
    & e^\mu_{(0)} = \left(\mathcal{E}/f,\dot{r},\mathcal{J}/r^2,0\right)\,,
    \\
    & e^\mu_{(3)} = \left(0,0,0,-\frac{1}{r}\right)\,,
\end{align}
where $\dot{r}$ is given by eq. \eqref{eq:dotr}. Note that the full expression for $\dot{r}$ including $\mathcal{O}(s)$ corrections avoids singularities in the definition of the tetrad at radial turning points.
Now the first and second legs are given as $e^\mu_{(1)} = \cos \psi\, \tilde{e}^\mu_{(1)}+\sin \psi \, \tilde{e}^\mu_{(2)}$ and $e^\mu_{(2)} = -\sin \psi\, \tilde{e}^\mu_{(1)}+\cos \psi \, \tilde{e}^\mu_{(2)}$, where
\begin{align}
    & \tilde{e}^\mu_{(1)} = \left(\frac{\dot{r} r \sqrt{h}}{\sqrt{f (\mathcal{J}^2 + r^2)}},\frac{\mathcal{E} r }{\sqrt{f h (\mathcal{J}^2 + r^2)}},0,0 \right)\,,
    \\
    & \tilde{e}^\mu_{(2)} = \left(\frac{\mathcal{E}\mathcal{J}}{f\sqrt{\mathcal{J}^2 + r^2}},\frac{\mathcal{J} \dot{r} }{r},\frac{\sqrt{\mathcal{J}^2 + r^2}}{r^2},0\right)\,,
\end{align}
and the precession angle $\psi(\tau)$ is obtained by integrating
\begin{align}
    & \dot{\psi} = \frac{\mathcal{E}\mathcal{J}}{\mathcal{J}^2 + r^2}\,.
\end{align}
The components of the spin with respect to the parallel-transported tetrad $e^\mu_{(A)}$ are constant at leading order. Additionally, since the spin magnitude $s \equiv \sqrt{s^\mu s_\mu}$ is conserved and $s_\mu e^\mu_{(3)} = s_\parallel + \mathcal{O}(s^2)$, the solution for the evolution of the spin vector can be expressed as
\begin{subequations}
\label{eq:sppsi}
\begin{align}
    & s^t = \sqrt{\frac{s^2 - s_\parallel^2}{f(\mathcal{J}^2 + r^2)}}\left(\frac{\mathcal{E \mathcal{J}} \cos \psi}{\sqrt{f}} + \dot{r} r \sin \psi\right) \,, 
    \\
    & s^r = \sqrt{s^2 - s_\parallel^2} \left( \frac{\mathcal{J} \dot{r} \cos \psi}{r} + \frac{\mathcal{E} r \sin \psi}{\sqrt{fh (\mathcal{J}^2 + r^2)}} \right)\,,
    \\
    & s^\vph = \frac{\sqrt{(s^2 - s_\parallel^2)(\mathcal{J}^2 + r^2)} \cos \psi}{r^2} \,,
    \\
    & s^\vth = -\frac{s_\parallel}{r}\,.
\end{align}
\end{subequations}
One can now also re-express $\delta \vth$ from eq. \eqref{eq:dvth} as 
\begin{align}
    \delta \vth = \frac{\sqrt{(s^2 - s_\parallel^2)(\mathcal{J}^2 + r^2)} \sin \psi}{\mathcal{J} r}\,. \label{eq:dvthpsi}
\end{align}

\noindent{\em Solution by quadrature.}--
The motion of the spinning test particle in any static, spherically symmetric metric is solved by quadrature as follows. First one integrates
\begin{align}
    & \tau(r) - \tau(r_0) = \pm\int_{r_0}^r \frac{ \di r'}{\sqrt{\mathcal{R}(r')}}\,, \label{eq:tauq}\\
    & \mathcal{R}(r) \equiv \frac{1}{h} \left( -1 + \frac{\mathcal{E}^2}{f} - \frac{\mathcal{J}^2}{r^2}\right) 
    + \frac{s_\parallel \mathcal{E} \mathcal{J} (2 f - r f')}{ (f h)^{3/2} r^2} \,,
\end{align}
and inverts this relation to obtain $r(\tau)$. This is then substituted into the $r$-parametrized solutions
\begin{align}
    & t(r) - t(r_0) = \pm\int_{r_0}^{r} \frac{\di r'}{\sqrt{\mathcal{R}(r')}}\left[\frac{\mathcal{E}}{f} + \frac{s_\parallel \mathcal{J} f'}{2 f r' \sqrt{fh}}\right] ,
    \\
    & \vph(r) - \vph(r_0) = \pm\int_{r_0}^{r} \frac{\di r'}{\sqrt{\mathcal{R}(r')}}\left[\frac{\mathcal{J}}{{r'}^2} + \frac{s_\parallel \mathcal{E} }{ {r'}^2 \sqrt{fh}}\right] , \label{eq:vphq}
    \\
    & \psi(r) - \psi(r_0) = \pm\int_{r_0}^{r} \frac{\di r'}{\sqrt{\mathcal{R}(r')}}\left[\frac{\mathcal{E}\mathcal{J}}{\mathcal{J}^2 + r'^2}\right] . \label{eq:psiq}
\end{align}
The aligned coordinate position $\vth = \pi/2 + \delta \vth$ and the spin components $s^\mu$ are then obtained by substituting the solutions for $r(\tau)$ and $\psi(\tau)$ into equations \eqref{eq:sppsi} and \eqref{eq:dvthpsi}.


\noindent{\em Schwarzschild space-time.}-- 
Let us demonstrate how to use this solution near a Schwarzschild black hole by setting $f = 1/h = 1-2M/r$. We focus on bound motion between radial turning points $r_1>r_2$ which are real roots of $\mathcal{R}(r)$. Following Darwin's treatment of geodesics \cite{darwin1959gravity}, we parametrize the motion by eccentricity $e$ and semi-latus rectum $p$ such that $r_1 = p/(1-e),\,r_2 = p/(1+e)$. The relation between energy, angular momentum and the orbital elements $e,p$ is obtained by inserting the geodesic relations from Darwin into the equation $\mathcal{R}(r) = 0$ and computing the spin corrections 
\begin{align}
    \begin{split}
    & \mathcal{E}^2 = \frac{(p - 2M)^2 - 4M^2 e^2}{p\left[p - M(3+e^2)\right]}
    \\
    & +s_\parallel \frac{(e^2 - 1)^2 M \sqrt{M p \left[p^2 - 4Mp - 4 M^2 (e^2 - 1)\right]}}{ p^2 \left[p - M(3+e^2)\right]^2} ,
    \end{split} \label{eq:E2pert}
    \\
    \begin{split}
    & \mathcal{J}^2 = \frac{M p^2}{p - M(3+e^2)} 
    \\
    &  - s_\parallel \frac{\left[2p - 3 M(3 + e^2)\right]\sqrt{M p \left[p^2 - 4Mp - 4 M^2 (e^2 - 1)\right]}}{ \left[p - M(3+e^2)\right]^2} .
    \end{split} \label{eq:J2pert}
\end{align}
Interestingly, these shifts exactly agree with those for a particle with spin fully aligned with orbital angular momentum when $s_\parallel = s$ (cf. Refs \citep{Drummond:2022xej,Mukherjee:2019jhd}). The bound motion exists for $e\in[0,1)$ and $p \in (p_{\rm c}(e),\infty)$, where the last stable orbits $p_{\rm c}(e)$ are determined by the vanishing of the Jacobian of the transform in eqs \eqref{eq:E2pert} and \eqref{eq:J2pert}, which yields (cf. the geodesic case in Ref. \cite{Cutler:1994pb})
\begin{align}
    p_{\rm c} = (6 + 2e)M + 2 s_\parallel\sqrt{\frac{2(1+e)}{3+e}} \,.
\end{align}
This generalises the results for innermost stable circular orbits of particles with aligned spin from Refs \cite{Suzuki:1997by,Favata:2010ic,Jefremov:2015gza,tsupko2016parameters} to fully generic motion.
The function $\mathcal{R}(r)$ can now be re-expressed as
\begin{align}
    &\mathcal{R} = \frac{1-\mathcal{E}^2}{r^4} (r_1 - r)(r-r_2)(r-r_3)(r-r_4)\,,\\
    &r_1>r_2>r_3>r_4\,,\; r_{1,2} = \frac{p}{1 \mp e}\,,\; r_4 = 0\,,
    \\
    \begin{split}
    & r_3 = \frac{2Mp}{p-4M} +\frac{2 s_\parallel \sqrt{M p \left[p^2 - 4Mp + 4M^2 (1-e^2)\right]}}{(p - 4M)^2}\,. \label{eq:r3}
    \end{split}
\end{align}

\begin{figure}
\begin{center}
\begin{minipage}{\linewidth}
    \begin{minipage}{0.27\textwidth}
        \includegraphics[width=\textwidth]{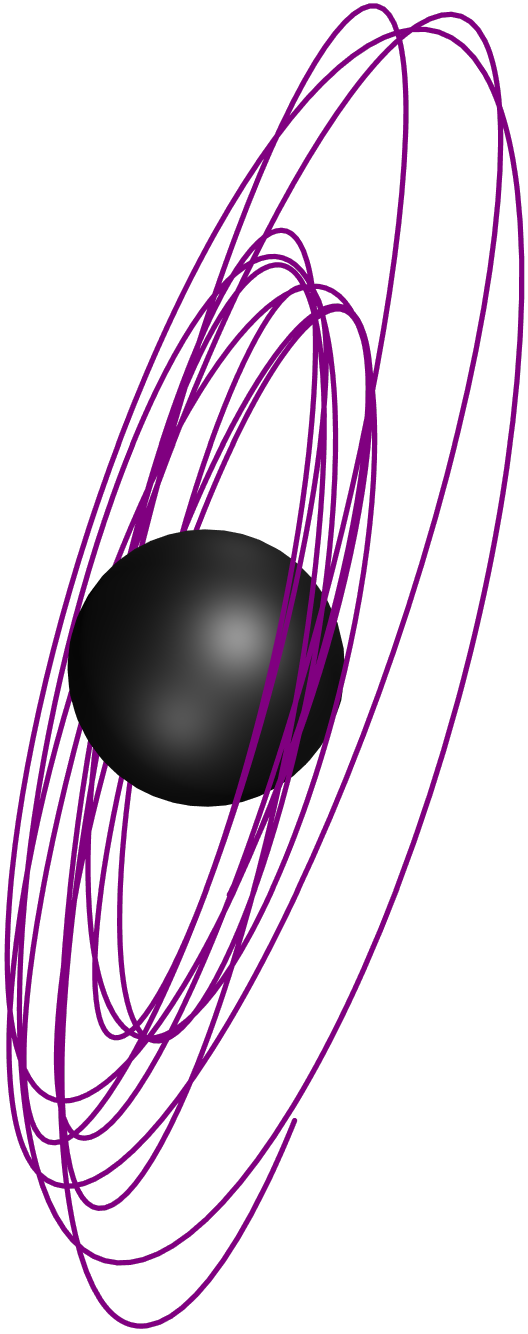}
    \end{minipage}
    $\;\,$
    \begin{minipage}{0.58\textwidth}
    \begin{center}
        \includegraphics[width=\textwidth]{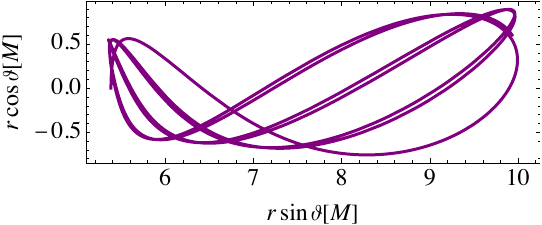} \\ 
        \includegraphics[width=0.82\textwidth]{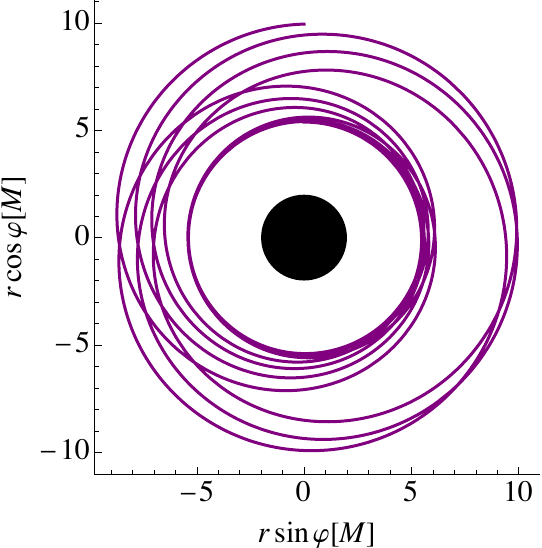} 
    \end{center}
    \end{minipage}
\end{minipage}
\caption{A generic inclined spinning particle orbit (left) and its description in angular-momentum-aligned coordinates (right) with $p=7M, e=0.3, s_\parallel = 0, s = 0.3M$.}
\label{fig:sporb}
\end{center} 
\end{figure}

\noindent{\em Expression as Jacobi elliptic integrals.}-- 
In Schwarzschild space-time, the most elegant parametrization of the motion is through Carter-Mino time \cite{carter1968,mino2003} $\lambda,\,\di \tau/\di \lambda = r^2$ so that
\begin{align}
    \begin{split}
    & \lambda(r) - \lambda(r_2) = \int_{r_2}^r\frac{\di r'}{\sqrt{R(r)}} 
    \\
    & \phantom{\lambda(r) - \lambda(r_2)} = \frac{2 F(\chi,k)}{\sqrt{(1 - \mathcal{E}^2)(r_1 -r_3)r_2}} \,,
    \end{split}
    \\
    & \sin \chi \equiv \sqrt{\frac{(r_1 - r_3)(r - r_2)}{(r_1 - r_2)(r - r_3)}}\,,
    \\
    & k^2 = \frac{(r_1 - r_2)r_3}{(r_1 - r_3)r_2}\,,
    \\
    & R(r) \equiv (1 - \mathcal{E}^2)(r_1-r)(r-r_2)(r-r_3)r\,,
\end{align}
where $F(\chi,k)$ is the elliptic integral of the first kind. All elliptic integrals $F,K,E,\Pi$ and their inverses ${\rm am},{\rm sn},{\rm cn}$ will be defined in the angle-modulus convention of Byrd \& Friedman \cite{byrd2013handbook} (see also Supplemental material).
The $\lambda(r)$ function can be inverted by the same arguments as given by van de Meent \cite{vandeMeent:2019cam} for corresponding formulas for Kerr geodesics to yield
\begin{align}
    & r(\lambda) = \frac{r_3(r_1 - r_2) \mathrm{sn}^2\left(\frac{K(k)}{\pi}  q^r, k\right) - r_2 (r_1 - r_3)}{(r_1 - r_2)\mathrm{sn}^2\left(\frac{K(k)}{\pi} q^r, k\right) - (r_1 - r_3)}\,, \label{eq:rlambda}\\ 
    & q^r \equiv \Upsilon^r \lambda + q^r_0\,,\\
    & \Upsilon^r \equiv \frac{\pi \sqrt{(1 - \mathcal{E}^2)(r_1 -r_3)r_2}}{2 K(k)}\,,
\end{align}
where $K(k)$ is the complete elliptic integral of the first kind and $q^r_0$ is an integration constant determined by initial conditions. Again, following closely the notation and approach of Ref. \cite{vandeMeent:2019cam} we express the other orbital variables as 
\begin{align}
    & t(\lambda) = q^t + \tilde{T}_{r}\left(\mathrm{am}\left(\frac{q^r}{\pi} K(k),k\right)\right) - \frac{\tilde{T}_{r}\left(\pi\right)}{2 \pi} q^r,\, \\
    & \vph(\lambda) = q^\vph + \tilde{\Phi}_{r}\left(\mathrm{am}\left(\frac{q^r}{\pi} K(k),k\right)\right) - \frac{\tilde{\Phi}_{r}\left(\pi\right)}{2 \pi} q^r, \label{eq:vphl}\\
    & \psi(\lambda) = q^\psi + \tilde{\Psi}_{r}\left(\mathrm{am}\left(\frac{q^r}{\pi} K(k),k\right)\right) - \frac{\tilde{\Psi}_{r}\left(\pi\right)}{2 \pi} q^r,\\
    & q^t \equiv \Upsilon^t \lambda + q^t_0,\,
    q^\vph \equiv \Upsilon^\vph \lambda + q^\vph_0,\,
    q^\psi \equiv \Upsilon^\psi \lambda + q^\psi_0,\,
\end{align}
where $q^{t,\vph,\psi}_0$ are again integration constants and
\begin{widetext}
\begin{align}
    \begin{split}
    & \tilde{T}_{r}(\chi) = \frac{\mathcal{E}}{\sqrt{(1 - \mathcal{E}^2)r_2(r_1-r_3)}} 
    \Bigg[ \frac{2M (r_1 - r_3)(r_2 - r_3) + r_3(r_3(r_2+r_3) - r_1(r_2 - r_3)) +2 \frac{\mathcal{J}}{\mathcal{E}} M s_\parallel}{r_3 - 2M} F(\chi,k)
    \\
    & \phantom{\tilde{T}_{r}(\chi) =}
    +(r_1+r_2+r_3+4M)(r_2 - r_3) \Pi\left(\chi,\frac{r_1 - r_2}{r_1 - r_3},k \right) + r_2 (r_1 - r_3) E(\chi,k)
    \\
    & \phantom{\tilde{T}_{r}(\chi) =}
    -\frac{2M (r_2 - r_3) \left(8 M^2 + \frac{\mathcal{J}}{\mathcal{E}} s_\parallel\right)}{(r_2 - 2M)(r_3 - 2M)} \Pi\left(\chi,\frac{(r_3-2M)(r_1 - r_2)}{(r_2 - 2M)(r_1 - r_3)},k \right)
    \\
    & \phantom{\tilde{T}_{r}(\chi) =}
    - \frac{(r_1 - r_2) \sin 2\chi \sqrt{r_2 (r_1 - r_3)\left[r_2(r_1 - r_3) -r_3(r_1 -r_2) \sin^2\! \chi\right]}}{r_1 + r_2 - 2 r_3 +(r_1 -r_2) \cos 2 \chi}
    \Bigg] \,, 
    \end{split}
    \\
    \begin{split}
     &  \tilde{\Phi}_{r}(\chi) = \frac{2( \mathcal{J} + s_\parallel \mathcal{E})}{\sqrt{(1 - \mathcal{E}^2)r_2(r_1-r_3)}} F(\chi,k) \,, 
    \end{split}
    \\
    \begin{split}
     &  \tilde{\Psi}_{r}(\chi) = \frac{2 \mathcal{E}\mathcal{J}}{\sqrt{(1 - \mathcal{E}^2)(r_1 - r_3)r_2} (\mathcal{J}^2 + r_3^2)} \Bigg[ 
     r_3^2 F(\chi,k) 
     +\frac{\mathcal{J}^2(r_2^2 - r_3^2)}{\mathcal{J}^2 + r_2^2} \mathfrak{Re}\left( \Pi \!\left(\chi,\frac{r_3-i \mathcal{J}}{r_2-i \mathcal{J}} \frac{r_1 - r_2}{r_1 - r_3},k\right)\right)
    \\
    & \phantom{\tilde{\Psi}_{r}(\chi) = }
    -\frac{\mathcal{J}(r_2 - r_3)(\mathcal{J}^2 - r_2 r_3)}{\mathcal{J}^2 + r_2^2} \mathfrak{Im}\left( \Pi\!\left(\chi,\frac{r_3-i \mathcal{J}}{r_2-i \mathcal{J}} \frac{r_1 - r_2}{r_1 - r_3},k\right)\right)
    \Bigg] \,, 
    \end{split} 
\end{align}
\begin{align}
\begin{split} 
    & \Upsilon^t = \frac{\Upsilon^r}{2\pi} \tilde{T}_{r}(\pi) = \frac{\mathcal{E}}{2 K(k)}
    \Bigg[ 
     \frac{2M (r_1 - r_3)(r_2 - r_3) + r_3(r_3(r_2+r_3) - r_1(r_2 - r_3))}{r_3 - 2M} K(k)
    \\
    & \phantom{\Upsilon^t = }
    +(r_1+r_2+r_3+4M)(r_2 - r_3) \Pi\!\left(\frac{r_1 - r_2}{r_1 - r_3},k \right) + r_2 (r_1 - r_3) E(k)
    \\
    & \phantom{\tilde{T}_{r}(\chi) =}
    -\frac{(r_2 - r_3)16 M^3}{(r_2 - 2M)(r_3 - 2M)} \Pi\!\left(\frac{(r_3-2M)(r_1 - r_2)}{(r_2 - 2M)(r_1 - r_3)},k \right) -\frac{\mathcal{J}s_\parallel}{\mathcal{E}}\bigg(1-\frac{r_1}{r_1 -2M} \Pi \!\bigg(\frac{2M(r_1-r_2)}{r_2(r_1-2M)}, k\bigg)\bigg)
    \Bigg]
    \,, 
\end{split}
    \\    
    \begin{split} 
    & \Upsilon^\vph = \frac{\Upsilon^r}{2\pi}\tilde{\Phi}_{r}(\pi) = \mathcal{J} + s_\parallel \mathcal{E} \,, 
    \end{split}
    \\
    \begin{split} 
    & \Upsilon^\psi =\frac{\Upsilon^r}{2\pi}\tilde{\Psi}_{r}(\pi) = \frac{\mathcal{E} \mathcal{J} r_3^2}{\mathcal{J}^2 + r_3^2} + \frac{\mathcal{E}\mathcal{J}^2 (r_2 - r_3)}{ K(k) (\mathcal{J}^2 + r_2^2)(\mathcal{J}^2 + r_3^2)} 
    \Bigg[\mathcal{J}(r_2+r_3)
     \mathfrak{Re}\left(\Pi \!\left(\frac{r_3-i \mathcal{J}}{r_2-i \mathcal{J}} \frac{r_1 - r_2}{r_1 - r_3},k \right)\right) 
     \\
     & \phantom{\Upsilon^\psi}
     - (\mathcal{J}^2 -r_2 r_3)
     \mathfrak{Im}\left(\Pi\!\left(\frac{r_3-i \mathcal{J}}{r_2-i \mathcal{J}} \frac{r_1 - r_2}{r_1 - r_3},k \right)\right)
    \Bigg] \,.  
    \end{split} \label{eq:Upsi}
\end{align}
\end{widetext}
This allows to plot fully explicit orbital solutions, as demonstrated in Fig. \ref{fig:sporb}. It should be noted that $\Upsilon^\psi, \tilde{\Psi}_r(\chi)$ can be evaluated at $s_\parallel = 0$ at leading order (with no spin correction to $r_3, \mathcal{E},\mathcal{J}$). On the other hand, $\Upsilon^{r,t,\vph},\tilde{T}_r, \tilde{\Phi}_r$ need to be explicitly expanded in $s_\parallel$ due to the spin corrections to $r_3, \mathcal{E},\mathcal{J}$ at fixed $p,e$ (see eq. \eqref{eq:E2pert},\eqref{eq:J2pert} and \eqref{eq:r3}). This procedure is straightforward, but the results are lengthy and thus relegated to the Supplemental material. From the ratios of these frequencies one can compute coordinate-time frequencies $\Omega^{r,\vph,\psi}\equiv \Upsilon^{r,\vph,\psi}/\Upsilon^t$ or the nodal and periastron precession rates, which is also discussed in the Supplement.

\noindent{\em Discussion and outlooks.}--  
The presented analytical solution for the motion of spinning test particles in Schwarzschild space-time can be taken ``as is'' to source gravitational-wave fluxes or to compute the gravitational self-force on the particle \cite{Warburton:2017sxk,Akcay:2019bvk,Mathews:2021rod,Skoupy:2021asz,Skoupy:2022adh,Skoupy:2023lih,Piovano:2020zin,Piovano:2021iwv,Piovano:2022ojl,Drummond:2023wqc}. Another useful output will be to extend the solution to compute action variables and scattering orbits in order to compare to Effective one-body models \cite{Buonanno:1998gg,Akcay:2020qrj,Albertini:2023aol,Khalil:2023kep} and other approaches to the relativistic two-body problem \cite{Blanchet:2012at,Bern:2020buy}. The formulas provided in the notebooks along with this Letter can also be straightforwardly implemented into existing code-bases such as the BHPToolkit \cite{BHPToolkit}, thus providing an immediate impact in gravitational-wave modeling. In particular, this solution is a springboard to understand the radiation-reaction on the fully generic motion of spinning particles in {\em Kerr} space-time \cite{Witzany:2019nml,Drummond:2022efc,Drummond:2023wqc}. Indeed, we have recently constructed semi-analytical orbits of spinning particles near Kerr black holes by using the Hamilton-Jacobi formalism presented in Ref. \cite{Witzany:2019nml}, and it is proving important to have independently derived analytical formulas for validation, even if just for the Schwarzschild case.  An important application will also be to use the quadratures in equations \eqref{eq:tauq}-\eqref{eq:psiq} to study deviations of the particle motion in modified gravities such as in Einstein-Gauss-Bonnet gravity \cite{Glavan:2019inb,Zhang:2020qew}, Horndeski and beyond-Horndeski theories \cite{Babichev:2017guv,Bakopoulos:2022csr}, or near boson stars~\cite{Kesden:2004qx,Macedo:2013jja,Diemer:2013zms,Brihaye:2014gua} and other exotic compact objects \cite{Cardoso:2019rvt}. 

However, we can already use our results to draw general conclusions about the detectability of secondary spin in large-mass-ratio inspirals onto spherically symmetric compact objects. Our solution demonstrates that time-averaged observables depend only on the aligned component of spin $s_\parallel$, independently of the compact-object model, and the contributions of the orthogonal component $ s_\perp, s_\perp^2 \equiv s^2-s_\parallel^2$ to any dynamical variable oscillate with the characteristic frequency $\Omega^\psi$ generically different from the orbital frequencies. The spin contributions to the post-1-adiabatic phase of the waveform depends only the spin shifts to quantities such as average orbital frequencies and the average energy and angular momentum fluxes, which are independent of $s_\perp$ (see also Ref. \cite{Skoupy:2023lih}). Furthermore, the $\Omega^\psi$-oscillation due to $s_\perp\lesssim m$ only appears in the $\mathcal{O}(m/M)$ sub-dominant corrections to the waveform amplitude~\cite{Flanagan:1997sx,Pound:2021qin}. 
As a result, waveform templates that neglect $s_\perp$ contributions will have a mismatch with models that include $s_\perp$ of order $\mathcal O(m^2/M^2)$. Hence, according to the Lindblom criterion, the waveforms that either include or neglect $s_\perp$ for EMRIs will be indistinguishable from each other in a matched filtering analysis unless the signal to noise ratio of the EMRI signal reaches the order of the large mass ratio $M/m \sim 10^4-10^6$ ~\cite{Lindblom:2008cm,schutz1991data}. This provides a key insight that will vastly simplify the treatment of spin in modified-gravity waveforms at large mass ratios.

Another consequence of this work is that the motion of spinning particles in any static spherically symmetric space-time is integrable to linear order in spin. In other words, if there is any chaos to be observed by the numerical integration of spinning particle motion \cite{Suzuki:1997by,Zelenka:2019nyp}, it is an $\mathcal{O}(s^2)$ effect that goes beyond the validity of the equations itself, which is in lines with the quantitative numerical scalings of resonances observed in Schwarzschild space-time \cite{Zelenka:2019nyp}. 

\noindent{\em Acknowledgements.}--  
We would like to thank Viktor Skoup{\'y} and Josh Mathews for pointing out typos in a draft of the paper. 
This work make use of the \textit{Wolfram Mathematica} package ``KerrGeodesics''~\cite{BHPToolkitKerrGeodesics}, which is a part of the BHPToolkit~\cite{BHPToolkit}.
VW is grateful for the support from the Charles University \textit{Primus} programme.
GAP aknowledges support from an Irish Research Council Fellowship under grant number GOIPD/2022/496.

\bibliography{literatura}
\newpage
\appendix
\onecolumngrid
\renewenvironment{widetext}{}{}
\section{ \large  Supplemental material for {\em Analytic solutions for the motion of spinning particles near spherically symmetric compact objects}}

\setcounter{equation}{0}
\renewcommand{\theequation}{S\arabic{equation}}

\subsection{Truncation of equations of motion at $\mathcal{O}(s)$}
\noindent The Mathisson-Papapetrou-Dixon equations of motion valid at quadrupolar order in a given background are  \citep{mathisson1937,papapetrou1951,dixon1964,dixon1970}
\begin{align}
    &\frac{\D P^\mu}{\di \tau} = - \frac{1}{2} R^\mu{}_{\nu\kappa \lambda}\dot{x}^\nu S^{\kappa \lambda} + \frac{1}{6} R_{\nu\kappa\lambda \gamma}{}^{;\mu} J^{\nu\kappa\lambda\gamma} + ...\,, \\
    &\frac{\D S^{\mu\nu}}{\di \tau} = P^\mu \dot{x}^\nu - P^\nu \dot{x}^\mu - \frac{4}{3} \left(R^\mu{}_{\kappa \lambda \gamma} J^{\nu \kappa \lambda \gamma} - R^\nu{}_{\kappa \lambda \gamma} J^{\mu \kappa \lambda \gamma} \right) + ...\,,
\end{align}
where $J^{\mu\nu\kappa\lambda}$ is the Dixon's quadrupole, which encodes the deformations of an extended body due to external tidal fields or its rotation \citep[see, e.g.][]{Bailey:1975fe,ehlers1977dynamics,Binnington:2009bb,steinhoff2012,steinhoff2016dynamical}. The three dots denote octupole and higher order multipoles which are not explicitly discussed here for the sake of simplicity.
The pole-dipole-quadrupole equations of motion are no longer universal because the quadrupole depends on the matter dynamics of the considered body.
Adiabatic and dynamical tidal quadrupoles \citep{Binnington:2009bb,steinhoff2016dynamical} turn out to be suppressed by the fifth power of the mass ratio in the large-mass ratio expansion, so we will not discuss them here.
However, $J^{\mu\nu\kappa\lambda}$ also includes a contribution to the quadrupole sourced by the rotation of the body. This term turns out to be second-order-in-mass-ratio as well as $\mathcal{O}(S^2)$ \citep{steinhoff2009}:
\begin{align}
    J^{\mu\nu\kappa\lambda} = \frac{3 m}{2 \mathcal{M}^3} c_{\rm ES^2} P^{[\mu} S^{\nu]}{}_\gamma S^{\gamma [\kappa} P^{\lambda]} \,,  
\end{align}
 where $[]$ denotes antisymmetrization of indices, $\mathcal{M}^2 \equiv -P_\mu P^\mu$, $m \equiv - P_\mu \dot{x}^\mu$, and $c_{ES^2}$ is a deformation coefficient that can be assumed to be approximately constant and $\propto m^3$ almost up to the mass-shedding limit \citep[see, e.g.,][]{Chakrabarti:2013tca}. The curvature is $R^\mu{}_{\nu\kappa\lambda}\sim 1/r_{\rm c}^2$, where $r_{\rm c}$ is the curvature length, and a covariant derivative adds an additional $\sim 1/r_{\rm c}$ factor. The Christoffel-symbol (parallel transport/geodesic) terms in the equations of motion scale as $\sim 1/r_{\rm c}$. The dipole terms thus scale with a relative factor $\sim S/r_{\rm c}$ as compared to the leading order in the equations of motion, and the quadrupole terms as $\sim S^2/r_{\rm c}^2$ relative to leading order. Similarly, higher multipoles contain higher powers of $S$ can be shown to scale with increasing powers of the ratio of spin and the curvature length.

The linear momentum is related to the four-velocity in the following manner under the Tulczyjew-Dixon condition $S^{\mu\nu}P_\nu = 0$ \citep{steinhoff2012}:
\begin{align}
    & \dot{x}^\mu = \mathcal{U}^\mu + \frac{2 S^{\mu \nu} S^{\kappa \lambda} R_{\kappa\lambda \nu \gamma} \mathcal{U}^\gamma}{4 \mathcal{M}^2 + S^{\gamma \delta} S^{\omega \chi} R_{\gamma \delta \omega \chi}}\,, \\
    & \mathcal{U}^\mu \equiv \frac{m}{\mathcal{M}^2} P^\mu - \frac{4}{3 \mathcal{M}^2} R^{[\mu}{}_{\nu\kappa\lambda}J^{\gamma]\nu\kappa\lambda} P_\gamma + \frac{1}{6 \mathcal{M}^2} S^{\mu\nu} R_{\kappa \lambda \chi \xi;\nu}J^{\kappa\lambda\chi\xi}\,.
\end{align}
One of the consequences of relation above is that $\mathcal{M}^2\dot{x}^\mu = m P^\mu + \mathcal{O}(S^2)$ so by projection into $\dot{x}_\mu$ we obtain $\mathcal{M}^2 = m^2 + \mathcal{O}(S^2)$ and both definitions of mass agree at leading order in spin.
By taking a covariant time derivative of the relation above and of $s^{\mu} \equiv \varepsilon^{\mu\nu\kappa\lambda}P_\nu S_{\kappa\lambda}/(2\mathcal{M}^2)$ we then obtain
\begin{align}
    &\frac{\D^2 x^\mu}{\di \tau^2} = - \frac{1}{4}{R}^\mu{}_{\nu\gamma \delta}\varepsilon^{\gamma\delta}{}_{\kappa\lambda} \dot{x}^\nu \dot{x}^\kappa s^\lambda + \mathcal{O}(s^2 R \nabla R)\,, \\
    &\frac{\D s^{\lambda}}{\di \tau} = 0 + \mathcal{O}(s^2 R^2) \,. 
\end{align}

In particular, the terms like $s^2 R \nabla R$ and $s^2 R^2$ have comparable contribution from the quadrupole $\propto c_{ES^2}$ coefficient as well from the non-trivial momentum-velocity relation in the Tulczyjew-Dixon supplementary spin condition. Nevertheless, all terms including the Riemann tensor and its derivatives scaled with the curvature length $r_{\rm c}$ in the same way. The Riemann tensor scales as $1/r_{\rm c}^2$ while the magnitude of the specific spin $s$ is set by the M{\o}ller radius $r_{\rm M}$, the minimal physical size of the rotating body \citep{moller1949definition}. Thus, for compact objects in the large mass ratio limit, the truncation of the equations of motion \eqref{eq:sacc} and \eqref{eq:strans} are controlled by powers of $r_{\rm M}/r_{\rm c}$. A multipole expansion is valid only when $r_{\rm M}/r_{\rm c} \ll 1$, a condition that is always satisfied for extreme mass-ratio binaries. This is because for bound motion near a black hole $r_c\sim \sqrt{r^3/M} \gtrsim M$ and thus also $1/r_c \lesssim 1/M$, and the M{\o}ller radius $r_{\rm M} \lesssim m$ for compact objects. As such, the equations of motion \eqref{eq:sacc} and \eqref{eq:strans} presented in the main text corresponds to the general form of the Mathisson-Papapetrou-Dixon equations of motion truncated at leading order in an expansion in $m/M$. Before adding the $\mathcal{O}(S^2)$ terms we would thus have to add leading-order gravitational self-force for self-consistence; the cross-interactions between self-force and spin would then be of the same importance as the $\mathcal{O}(S^2)$ terms in the equations of motion, which is beyond the scope of our Letter.


\subsection{Aligning the coordinate frame}

\noindent Given the components of the angular-momentum vector $\vec{\mathcal J}$ computed from equations \eqref{eq:Jx}-\eqref{eq:Jz} in the main text and some initial conditions, the inclination angle $\iota \in (0,\pi)$ of $\vec{\mathcal J}$ with respect to the $z$ direction is given by
\begin{align}
    & \cos \iota = \frac{\vec{\mathcal J}\cdot \hat{z}}{\mathcal J} = \frac{\mathcal{J}_z}{\mathcal{J}}\,,\\
    & \mathcal{J} \equiv \sqrt{\mathcal{J}_x^2+\mathcal{J}_y^2+\mathcal{J}_z^2}\,,
\end{align}
where $\hat{z}$ is the unit vector in the $z$ direction.
Now the aligned frame is obtained by rotating by $\iota$ around the axis $\vec{n}$ given as
\begin{align}
    & \vec{n} = \frac{\vec{\mathcal{J}}\times \hat{z}}{|\vec{\mathcal{J}}\times \hat{z}|} = (-\cos \xi,\sin \xi,0)\,, \\
    & \xi \equiv \arctan \frac{\mathcal{J}_x}{\mathcal{J}_y}\,,
\end{align}
where $\xi$ is related to the longitude of the ascending node known from classical celestial mechanics \cite{morbidelli2002modern}. The angles $\iota,\xi$ can be used to parametrize the angular-momentum vector as
\begin{align}
    \vec{\mathcal J} = \mathcal{J} (\sin \xi \sin \iota, \cos \xi \sin \iota,\cos \iota) \,.
\end{align}
The transformation of any vector into the aligned coordinates then reads
\begin{align}
    \vec{v}' = \vec{n}(\vec{n}\cdot\vec{v}) + \cos\iota (\vec{n}\times \vec{v})\times \vec{n} - \sin \iota (\vec{n} \times \vec{v})\,.
\end{align}
It is now easy to show that this map yields $\vec{\mathcal J}' = (0,0,\mathcal{J}_{z'} = \mathcal{J})$. Note that the transformation also flips counter-rotating coordinate systems ($\mathcal{J}_z <0$) into co-rotating ones ($\mathcal{J}_{z'} >0$), which corresponds to $\iota>\pi/2$. The inverse map is simply obtained by switching the sign on the angle of rotation $\iota$
\begin{align}
    \vec{v} = \vec{n}(\vec{n}\cdot\vec{v}') + \cos\iota (\vec{n}\times \vec{v}')\times \vec{n} + \sin \iota (\vec{n} \times \vec{v}')\,.
\end{align}
 From the transformation of the unit position vector $\hat{r} = (\sin \phi \sin \theta, \cos \phi \sin \theta, \cos \theta)$ and the aligned vector $\hat{r}' = (\sin \vph \sin \vth, \cos \vph \sin \vth, \cos \vth)$ one can easily read off the aligned coordinate system and its inverse transformation. Specifically, one can take the $t,r,\vph,\vth = \pi/2+\delta \vth$ solution presented in the main text of the paper and map it back into the original coordinates ($t,r$ are unchanged) 
 \begin{widetext}
 \begin{align}
     & \cos \theta =  \cos \vth \cos \iota - \sin \vth  \sin \iota \cos(\xi - \vph)\,, \label{eq:thinv}\\
     & \tan \phi = \frac{\cos \vth \sin \iota \sin \xi + \sin \vth \left[\cos^2\! \frac{\iota}{2} \sin \vph - \sin^2\! \frac{\iota}{2} \sin (2 \xi - \vph)\right]}{\cos \vth \sin \iota \cos \xi + \sin \vth \left[\cos^2\! \frac{\iota}{2} \cos \vph - \sin^2\! \frac{\iota}{2} \cos (2 \xi - \vph)\right]}\,. \label{eq:phinv}
\end{align}  
 \end{widetext}

\subsection{Expressing $\delta \vth$}

\noindent Now let us express $\vth(\tau) = \pi/2+\delta \vth(\tau)$ from $\mathcal{J}_x = \mathcal{J}_y =0$. First, let us examine the $\mathcal{J}_x =0$ expression
\begin{align}
\begin{split}
    &0 = \cos \vph \left[r^2 \delta \vth  \dot{\vph} + \frac{f h (s^t \dot{r} - s^r \dot{t})}{\sqrt{fh}}\right] 
     + \sin \vph \left[-r^2 \delta \dot{\vth}  + \frac{f (s^\vph \dot{t} - s^t \dot{\vph})}{\sqrt{fh}}\right] \,. \label{eq:Jx0}
\end{split}
\end{align}
The $\mathcal{J}_y$ expression is identical up to $\sin \vph \leftrightarrow \cos \vph$. This means that the expressions in the square brackets in eq. \eqref{eq:Jx0} must vanish independently. This may seem like an overconstrained system since both $\delta \vth$ and $\delta \dot{\vth}$ fulfill apparently independent conditions. However, one can see that by taking the expression for $\delta \vth$ from eq. \eqref{eq:dvth}, taking a time-derivative, and substituting equations of motion, the two equations yield the same information and are mutually consistent.

Once we have expressed $\delta \vth$ using the solution for $r,\psi$ and the solution for $\vph$, all presented in the main text (cf. eqs \eqref{eq:dvthpsi}-\eqref{eq:psiq}), we obtain the explicit evolution of the generic coordinates $\theta,\phi$ by linearizing the formulas \eqref{eq:thinv} and \eqref{eq:phinv} to obtain
\begin{align}
\begin{split}
    & \theta = \pi -\arccos\left[\cos(\xi-\vph) \sin \iota\right] 
    + \frac{\delta \vth \cos \iota}{\sqrt{1 - \cos^2(\xi - \vph) \sin^2\! \iota}} \,,
\end{split}\\
\begin{split}
    & \phi = \arctan\left[\frac{\cos^2\! \frac{\iota}{2} \sin \vph - \sin^2\! \frac{\iota}{2}\sin(2\xi - \vph)}{\cos^2\! \frac{\iota}{2} \cos \vph - \sin^2\! \frac{\iota}{2}\cos (2\xi - \vph)}\right]
   -\frac{4 \delta \vth \sin \iota \sin(\xi - \vph)}{3 + \cos 2\iota - 2 \cos(2\xi - 2\vph) \sin^2\! \iota  }
   \,.
\end{split}
\end{align}

\subsection{Jacobi elliptic integrals}
\noindent The convention of Byrd \& Friedman \cite{byrd2013handbook} for incomplete Jacobi elliptic integrals of the first, second and third kinds $F,E,\Pi$ is as follows
\begin{align}
    & F(\chi,k) \equiv \int_0^\chi \frac{\di \xi}{\sqrt{1 - k^2 \sin^2\! \xi}} \,,\\
    & E(\chi,k) \equiv \int_0^\chi \sqrt{1 - k^2 \sin^2\! \xi} \,\di \xi \,,\\
    & \Pi(\chi,n,k) \equiv \int_0^\chi \frac{\di \xi}{(1 - n \sin^2\! \xi)\sqrt{1 - k^2 \sin^2\! \xi}} \,.\\
\end{align}
The complete elliptic integrals are then defined by $K(k) \equiv F(\pi/2,k),\,E(k) \equiv E(\pi/2,k),\,\Pi(n,k) \equiv \Pi(\pi/2,n,k)$. The Jacobi amplitude is the inverse of the elliptic integral of the first kind $F({\rm am}(u,k),k) \equiv u$ and we also use ${\rm sn}(u,k) \equiv \sin({\rm am}(u,k)),\,{\rm cn}(u,k) \equiv \cos({\rm am}(u,k))$.

This convention should be contrasted with the convention of \textit{Maplesoft Maple}, which uses $\sin (\xi)$ in the first argument of incomplete integrals, or \textit{Wolfram Mathematica}, which uses $k^2$ in the last argument and switches the order of the first two arguments in the incomplete elliptic integral of the third kind $\Pi$.

\subsection{Linear shift to the orbital frequencies}
\noindent We write the linear correction to the Mino frequencies as 
\begin{align}
    \Upsilon^{r,t,\vph} = \Upsilon^{r,t,\vph}_{\rm (g)} + s_\parallel \delta \Upsilon^{r,t,\vph}\,,
\end{align}
where $\Upsilon^{r,t,\vph}_{\rm (g)}$ are the known  expressions for geodesic Mino frequencies \cite{fujita2009,vandeMeent:2019cam}, and we expand at fixed $p,e$. We first define the various arguments appearing in the elliptic integrals
\begin{align}
    & k^2_0\equiv  \frac{4M e}{p+2M(e -3)}  \,, \\ 
    & \alpha_0 \equiv  \frac{2e (p -4M)}{(1+e)\left[p - 2M(3-e)\right]} \,,\\
    & \beta_0 \equiv  \frac{16 M^2 e}{\left[p - 2M(3-e)\right]\left[p - 2M(1+e)\right]} \,,\\
    & \gamma_0 \equiv \frac{2Me \left[p - 4M +2i \sqrt{p - 3M(1+e^2)} \right]}{\left[p - 2M(3-e)\right] \left[M(1 + e) + i\sqrt{p - 3M(1+e^2)}\right]}.
\end{align}
The frequency shifts are then
\begin{widetext}
 \begin{align}
    & \delta \Upsilon^r = \frac{\pi M\sqrt{(p+2M(e -3))((p-2M)^2 -4M^2 e^2)}\big[(p-2M(e +3)) K(k_0) - (p-M(e^2 +3))E(k_0) \big] }{4(p -2M( e+3))(p- M(e^2 +3))^{3/2}K(k_0)^2} \,,\\
    & \delta \Upsilon^t = \frac{p \sqrt{M}}{2\sqrt{p- M(e^2 +3)}}\bigg( \frac{p \Pi \big(\frac{4 e}{(p-2M(1-e)}, k_0\big)}{(p-2M(1-e))K(k_0)} -1 \bigg) -\sqrt{M}\frac{A(p,e)E(k_0)^2 +B(p,e) E(k_0) + C(p,e)}{8M(p-4M)^3(p- M(e^2 +3))^{3/2}K(k_0)^2}   \,,\\
    &\delta \Upsilon^\vph = \frac{(3+e^2) M\sqrt{p[(p-2M)^2-4 M^2 e^2]}}{2p (p-M(e^2+3))^{3/2}} \,,
 \end{align}
where
\begin{align}
 &A(p,e) = \frac{2p(p-M(e^2+3))(p-2M(e+1))(p-4M)^2(p+2M(e-3))(p-2M(1-e))}{(1-e^2)(p-2M(e+3))} \ , \\
 \begin{split}
 &B(p,e) = -\frac{2(p-4M)}{(1-e)(1+e)^2} \bigg[(1 + e)p(p+2M(e-3)) K(k_0) \Big(e^4 M^2 (p+4M)-2e^2M(p-3M)(p+4M)-28M^3+ \\
 & \phantom{B(p,e) =} + 33M^2 p+ 2p^2(p-7M)\Big)+2(p-M(e^2+3))(p-2M(1-e))\bigg((1-e)(1+e)^2M(p-4M)^2 \Pi(\gamma_0, k_0) + \\
 & \phantom{B(p,e) =} +(p -2M(1+e))(8M^2+Mp-p^2+e^2M(3 p -8M)) \Pi \!\left(\alpha_0, k_0\right) \!\bigg) \bigg] \ ,
 \end{split}
\end{align}
\begin{align}
\begin{split}
  &C(p,e)  = \frac{K(k_0)}{(1-e)(1+e)^2(p-2M(1 +e))} \bigg[(p-2M(e +3))\bigg(  \\
  &\phantom{C(p,e) =}(1-e)(1+e)^2(p-4M)^4 (-4M^2-(p-3M)p+e^2 M(p+4M))\Pi(\gamma_0, k_0)+ \\
  & \phantom{C(p,e) =}+4(p-2M(1+e)) \Pi \! \left(\alpha_0, k_0\right)\! \bigg(
  3M^3e^6p^2+e^4M^2(p(-6p^2+3Mp+80M^2)-128M^3)-128M^5+  \\
  & \phantom{C(p,e) =}+144M^4p-p^4(p-6M)-2M^2p^3-55M^3p^2+Me^2(6p^4-40Mp^3+113M^2p^2-224M^3p+256M^4)\bigg)\!\bigg) +  \\
  &\phantom{C(p,e) =}+(1 + e)(p-2M(1+e))pK(k_0)\bigg(24e^6M^4p+e^4(M(p-10M)p^3+184M^4p-32M^3p^2-128M^5)- \\
  &\phantom{C(p,e) =}-e^2(952M^4p-288M^3p^2+76M^2p^3+(p-16M)p^4-1280M^5)+3(p-19M)p^4+374M^3p^3- \\
  &\phantom{C(p,e) =}-1152M^3p^2+1768M^4p-1152M^5\bigg)\bigg] \ .
\end{split}
\end{align}
The radial shift $ \delta \Upsilon^r$ agrees with the numerical integration of eq. (B26) in Ref. \cite{Drummond:2022xej}. We checked the azimuthal shift $\delta \Upsilon^\vph$ and the shift $\delta \Upsilon^t$ against the frequency shifts obtained from the Hamilton-Jacobi formalism \cite{Witzany:2019nml}.
The precession frequency written explicitly to leading order in terms of $p,e$ reads
\begin{equation}
\Upsilon^\psi = \frac{\sqrt{M p}}{\sqrt{(p-2M)^2-4 M^2 e^2}K(k_0)}
    \Bigg[4 M K(k_0)+(p-2M(e +3))\bigg(
     \frac{\sqrt{M}(1-e)}{\sqrt{p-M(e^2+3)}}\mathfrak{Im}\left[\Pi(\gamma_0,k_0)\right]
     +\mathfrak{Re}\left[\Pi(\gamma_0,k_0)\right] \bigg)
    \Bigg] , 
\end{equation}
which agrees with eq. (64) of Ref. \cite{vandeMeent:2019cam}. 

Using the previous expressions, it is easy to compute the nodal and periastron precession $\nu_{\rm nodal}$ and $ \nu_{\rm peri}$, respectively, which are defined as
\begin{align}
   \nu_{\rm nodal} &= 2\pi\bigg(\frac{\Upsilon^\vph}{\Upsilon^\psi} - 1\bigg)  \,, \quad
   \nu_{\rm peri} = 2\pi\bigg(\frac{\Upsilon^\vph}{\Upsilon^r} - 1\bigg)  \ ,
\end{align}
The average polar fundamental frequency is $\Upsilon^\psi$,  as can be seen from eq. \eqref{eq:dvthpsi}. 
This leads to the explicit formulas
\begin{align}
   \nu_{\rm nodal} &=  2\pi\left(\frac{1}{\mathcal{K}(p,e)}\sqrt{\frac{p(p-2M(1-e))(p-2M(1+e))}{p-M(e^2+3)}}K(k_0) -1\right)  \,, 
\end{align}
where
\begin{align}
   \mathcal{K}(p,e) &=4 M K(k_0)+(p-2M(e +3))\left(
     \frac{\sqrt{M}(1-e)}{\sqrt{p-M(e^2+3)}}\mathfrak{Im}\left[\Pi(\gamma_0,k_0)\right]
     +\mathfrak{Re}\left[\Pi(\gamma_0,k_0)\right] \right) \, .
\end{align}
\end{widetext}
We stress that $\nu_{\rm nodal}$ is  evaluated at  leading $s_\parallel =s=0$ order, since the precession itself is an $\mathcal{O}(s)$ effect. On the other hand, as can be seen from eq. \eqref{eq:vphl}, $\vph$ will evolve by exactly the same amount between two pericenter passages and no averaging needs to be invoked. The non-averaged periastron precession is then given by
\begin{align}
   \nu_{\rm peri} = \nu^{\rm g}_{\rm peri} + s_\parallel  \delta\nu_{\rm peri}  \ ,
\end{align}
where
\begin{align}
   \nu^{\rm g}_{\rm peri} &= \frac{4p K(k_0)}{\sqrt{p(p+2M(e-3))}} -2\pi\ \, , \\
    \begin{split}
    \delta\nu_{\rm peri} &= 2\sqrt{\frac{(p-2M)^2-4M^2e^2}{p(p+2M(e-3))}}\bigg(\frac{E(k_0)}{p-2M(e+3)}
    -\frac{K(k_0)}{p}\bigg) \, .
    \end{split}
\end{align}

\subsection{Linear shift to the orbital motion}

\noindent The expansion of the orbital motion in spin is non-unique. We choose to expand at fixed radial turning points (parametrized by $p,e$) and fixed angle variables (homogeneous orbital phases) $q^t,q^r,q^\vph,q^\psi$. This is because the angle variables smoothly parametrize the orbital motion and avoid singularities at turning points associated, e.g.,  with expansions at fixed velocities and coordinates $u^\mu,x^\mu$. Furthermore, this separates the ``oscillating'' part of the spin corrections from the secular/averaged effect of the spin expressed by the frequency shifts $\delta \Upsilon^{t,r,\vph}$. By this method we obtain $x = x_{(\rm g)} + s_\parallel \delta x$ for $x = t,r,\vph$. On the other hand, $\psi$ is consistent to compute only to leading order, since it appears only in $\mathcal{O}(s)$ observables. The resulting expressions read
\begin{widetext}
    \begin{align}
        & r_{(0)} = \frac{p\left[p -2M(3-e) - 4M e \sin^2\! \chi_0\right]}{p - 2M(3-e^2) + e(p-4M) \cos 2\chi_0}\,,\\
    \begin{split}
       & t_{(0)} = q^t + \sqrt{\frac{(p - 2M)^2 - 4 e^2}{M\left[p -2M(3-e)\right]}} \Bigg[ 
        \frac{p\left[p^2-14M p + 2M e^2 (p-2M) +36 M^2\right]\left(F(\chi_0,k_0) - \frac{q^r}{\pi} K(k_0)\right)}
        {(e^2 -1)(p-4M)^2} 
        \\ & \phantom{t_{(0)} =} 
        + \frac{2\left[p -2M(3-e)\right]\left[p^2 - Mp (1 + 3 e^2)+8M^2 (e^2 - 1)\right] \left(\Pi(\chi_0,\alpha_0,k_0) - \frac{q^r}{\pi}\Pi(\alpha_0,k_0) \right)}{(1-e^2)(1+e)(p-4M)^2}
        \\ & \phantom{t_{(0)} =} 
        - \frac{2M\left[p -2M(3-e)\right]\left(\Pi(\chi_0,\beta_0,k_0) - \frac{q^r}{\pi}\Pi(\beta_0,k_0) \right)}{p - 2M(1-e)}
        +\frac{M p\left[p -2M(3-e)\right] \left(E(\chi_0,k_0) - \frac{q^r}{\pi} E(k_0)\right)}{(1-e^2)(p-4M)^2}
        \\ & \phantom{t_{(0)} =} 
        - \frac{e p \sin 2 \chi_0 \sqrt{\left[p -2M(3-e)\right]\left[p -2M(3-e \cos 2 \chi_0)\right]}}{(1-e^2)\left[p - 2M(3-e^2) + e(p-4M) \cos 2\chi_0\right]}
        \Bigg]
        \,,
    \end{split} \\
    \begin{split}
       & \vph_{(0)} = q^\vph + 2\sqrt{\frac{p}{p - 2M(3-e)}} \left(F(\chi_0,k_0) - \frac{q^r}{\pi} K(k_0)\right)
        \,,
    \end{split} \\
    \begin{split}
       & \psi = q^\psi + \frac{2}{\sqrt{\left[(p- 2M)^2 -4M^2 e^2\right] \left[p - 2M(3-e)\right]}}  \Bigg[ 
        4M\sqrt{p - M(3+e^2)}\left(F(\chi_0,k_0) -  \frac{q^r}{\pi} K(k_0)\right)
        \\ & \phantom{\psi =} 
        + \sqrt{M}(1-e)\left[p - 2M(3+e)\right] \mathfrak{Im}\left( \Pi(\chi_0,\gamma_0,k_0) - \frac{q^r}{\pi}\Pi(\gamma_0,k_0)\right)
        \\ & \phantom{\psi =} 
        + \sqrt{p - M(3+e^2)}\left[p - 2M(3+e)\right] \mathfrak{Re}\left( \Pi(\chi_0,\gamma_0,k_0) - \frac{q^r}{\pi}\Pi(\gamma_0,k_0)\right)
        \Bigg]
        \,,
    \end{split} \\
        & \delta r = 
        -\frac{
        e \sqrt{p \left[(p - 2M)^2 - 4 e^2\right]} \left[p -2M(3-e)\right] \sin 2\chi_0 \sqrt{1 - k_0^2 \sin^2 \! \chi_0 } \left( E(\chi_0,k_0) - \frac{q^r}{\pi} E(k_0)\right)
        }{
        \left[p - 2M(3-e^2) + e(p-4M) \cos 2\chi_0 \right]^2
        } \,, \\
& \chi_0 \equiv \mathrm{am}\left(\frac{q^r}{\pi} K(k_0),k_0\right) \,.
\end{align}
\end{widetext}
The $\delta t, \delta \vph$ functions are too long to be typed here explicitly, but they can be found in machine-readable form in the accompanying file \textit{SchwarzschildSpinAnalytic.m}. 

It should be noted that if one is only interested in the approximate orbit, it is computationally more efficient to use the non-expanded formulas \eqref{eq:rlambda}-\eqref{eq:Upsi} presented in the main text. On the other hand, in a two-timescale expansion of a gravitational-wave inspiral it is convenient to separate the various pieces contributing to the motion separately, which is the main reason to provide these formulas.

Finally, the expanded results can now be substituted to obtain $\delta \vth(q^r,q^\psi,p,e)$ as
\begin{align}
    \delta \vth = \frac{\sqrt{M(s^2 - s_\parallel^2)\left[M p^2+ (p-M(3+e^2)) r^2_{(0)}\right]} \sin \psi}{M p r_{(0)}}\,. \label{eq:dvthpsiexpl}
\end{align}

\end{document}